# Oligarchic planetesimal accretion and giant planet formation II
# *(Research Note)*

A. Fortier[1,2,*], O. G. Benvenuto[1,2,**], and A. Brunini[1,2,***]

[1] Facultad de Ciencias Astronómicas y Geofísicas, Universidad Nacional de La Plata, Paseo del Bosque s/n (B1900FWA) La Plata, Argentina
e-mail: afortier@fcaglp.unlp.edu.ar

[2] Instituto de Astrofísica de La Plata, IALP, CCT-CONICET-UNLP, Argentina

**ABSTRACT**

*Aims.* The equation of state calculated by Saumon and collaborators has been adopted in most core–accretion simulations of giant–planet formation performed to date. Since some minor errors have been found in their original paper, we present revised simulations of giant–planet formation that considers a corrected equation of state.
*Methods.* We employ the same code as Fortier and collaborators in repeating our previous simulations of the formation of Jupiter.
*Results.* Although the general conclusions of Fortier and collaborators remain valid, we obtain significantly lower core masses and shorter formation times in all cases considered.
*Conclusions.* The minor errors in the previously published equation of state have been shown to affect directly the adiabatic gradient and the specific heat, causing an overestimation of both the core masses and formation times.

## 1. Introduction

It is widely accepted that the most likely mechanism of giant–planet formation is the nucleated instability hypothesis (for a review on this topic see, e.g., Lissauer & Stevenson 2007). According to this model, planetesimals within a young protoplanetary disc collide with each other, eventually in some cases coagulating into more massive bodies. If these "planetary embryos" appear in the disc while the nebular gas is still present, they may be able to bind gravitationally large amounts of gas from the surrounding nebula to become, in the end, giant planets. Therefore, the entire formation process must occur before the dissipation of the protoplanetary disc. Circumstellar discs are mainly observed around very young stars (ages $\lesssim 10^7$ years, Hillenbrand 2005). This implies that the formation timescale of giant planets is a tight constraint. This limit has challenged nucleated instability models because it has been very difficult to find formation times compatible with the estimations of the disc lifetime (Pollack et al. 1996, hereafter P96; Thommes at al. 2003; Hubickyj et al. 2005; Alibert et al. 2005). For studies in which gas and solids accretion rates are calculated interactively, the solid growth regime has been generally overestimated, leading to core formation times that are a negligible fraction of the total formation time. Therefore, these means of estimating formation timescales are based mainly on the gas–accretion phase.

On the other hand, *N*-body simulations that study the growth of solid embryos in detail show that even bodies of several hundredths of the Earth mass can perturb significantly the neighbouring planetesimals, increasing their relative velocities and reducing the growth rate of the embryo itself (Ida & Makino 1993). This regime is known as "oligarchic growth" since only the most massive bodies in the disc can continue with the accretion (Kokubo & Ida 1998). The cores of giant planets are solid bodies that could be readily differentiated from the swarm of planetesimals and, afterwards, grew according to the oligarchic regime.

In Fortier et al. (2007), hereafter Paper 1, we presented results of self–consistent giant–planet formation calculations adopting the oligarchic regime for the growth of the core. The aim of our study was to investigate the nucleated instability model with a more realistic description of the solid's accretion rate. After our results were published, we discovered that there were some typos in the original article of the equation of state (EOS). Saumon et al. (1995), hereafter SCVH, calculated the EOS for pure hydrogen and helium. To approximate a solution for a mixture of these gases, they provided an interpolation formula for the entropy of mixing. However, in their paper, some of the equations given to calculate the resulting entropy contained minor errors that for our calculations, affected both the formation times and the final mass of the cores. Although, qualitatively, the conclusions of our previous article still hold, the new results differ quantitatively. Therefore, we decided to repeat all our simulations and present our updated results, especially because the supermassive cores in our former simulations had been found

---

*Send offprint requests to*: A. Fortier

* Fellow of the Consejo Nacional de Investigaciones Científicas y Técnicas (CONICET), Argentina.

** Member of the Carrera del Investigador Científico, Comisión de Investigaciones Científicas de la Provincia de Buenos Aires, Argentina. E-mail: obenvenu@fcaglp.unlp.edu.ar

*** Member of the Carrera del Investigador Científico, CONICET, Argentina. E-mail: abrunini@fcaglp.unlp.edu.ar

to be artificial. We also include additional calculations for the formation of a Jupiter–like object.

## 2. The model in brief

We summarise the main properties of our model, which are the same as in Paper 1. We developed a numerical code to calculate the formation of giant planets in the framework of the core–instability hypothesis (Mizuno 1980, P96). We considered that the protoplanet is immersed in a disc, whose temperature and density profile correspond to the standard Minimum Mass Solar Nebula (MMSN) of Hayashi (1981). We assumed a oligarchic growth regime for the core (Ida & Makino 1993) from the beginning of the calculations[1]. The gas accretion was coupled in an interactive fashion with the accretion of planetesimals. The external radius of the protoplanet was defined to be $R = \min[R_{\text{acc}}, R_H]$, where $R_{\text{acc}}$ is the accretion radius and $R_H$ is the Hill radius. The full differential equations of stellar evolution were solved with an adaptation of a Henyey–type finite difference code. Radiative and convective transport are considered by applying the Schwarzschild criterion for the onset of convection. The adiabatic gradient was adopted as the temperature gradient in the latter case. For the grain opacities, we used the tables of Pollack et al. (1985). For temperatures above $10^3$ K, we considered Alexander & Ferguson (1994) molecular opacities, which are available for $T \leq 10^4$ K, and for higher temperatures we considered the opacities by Rogers & Iglesias (1992). The EOS was that of SCVH.

SCVH calculated EOS tables for pure hydrogen and helium and suggested that for a mixture, the EOS can be approximated by interpolating the data in both tables. The equations required to complete the interpolations are given in Sect. 6 of SCVH. Unfortunately, there were some typos in their equations, and here we rewrite the equations in their correct form.

According to SCVH, the total entropy $S$ of the mixture can be approximated to be

$$S(P,T) = (1-Y)S^H(P,T) + Y S^{\text{He}}(P,T) + S_{\text{mix}}(P,T) \quad (2)$$

where $Y$ is the mass fraction of helium, $S^H(P,T)$ ($S^{\text{He}}(P,T)$) is the hydrogen (helium) entropy for a given pressure and temperature, and $S_{\text{mix}}$ is the ideal entropy of mixing, calculated to be

$$S_{\text{mix}} = k_B \frac{1-Y}{m_H} \frac{2}{1 + X_H + 3X_{H_2}} \times \{\ln(1+\beta\gamma)$$
$$- X_e^H \ln(1+\delta) + \beta\gamma[\ln(1+1/\beta\gamma)$$
$$- X_e^{\text{He}} \ln(1+1/\delta)]\} \quad (3)$$

where $\beta$, $\gamma$, and $\delta$ are

$$\beta = \frac{m_H}{m_{\text{He}}} \frac{Y}{1-Y}, \quad (4)$$

$$\gamma = \frac{3}{2} \frac{(1 + X_H + 3X_{H_2})}{(1 + 2X_{\text{He}} + X_{\text{He}^+})}, \quad (5)$$

$$\delta = \frac{2}{3} \frac{(2 - 2X_{\text{He}} - X_{\text{He}^+})}{(1 - X_{H_2} - X_H)} \beta\gamma, \quad (6)$$

---
[1] Eq. (5) of paper 1 should read

$$v_{\text{rel}} \simeq \sqrt{\langle e_m^2 \rangle + \langle i_m^2 \rangle} v_k. \quad (1)$$

respectively and $X_A$ is the concentration of particles A. The logarithmic derivatives of the entropy with respect to the temperature and the pressure ($S_T$, $S_P$) are

$$S_T = \left.\frac{\partial \log S}{\partial \log T}\right|_P = (1-Y)\frac{S^H}{S}S_T^H + Y\frac{S^{\text{He}}}{S}S_T^{\text{He}} + \frac{S_{\text{mix}}}{S}\left.\frac{\partial \log S_{\text{mix}}}{\partial \log T}\right|_P, \quad (7)$$

$$S_P = \left.\frac{\partial \log S}{\partial \log P}\right|_T = (1-Y)\frac{S^H}{S}S_P^H + Y\frac{S^{\text{He}}}{S}S_P^{\text{He}} + \frac{S_{\text{mix}}}{S}\left.\frac{\partial \log S_{\text{mix}}}{\partial \log P}\right|_T. \quad (8)$$

In the original paper, the typos were in Eq. (56) of SCVH (here Eq. (6)), where the factor $\frac{2}{3}$ was instead written $\frac{3}{2}$, and in Eqs. (45), and (46) (here Eqs. (7), and (8)), where the quotients $\frac{S^H}{S}$ and $\frac{S^{\text{He}}}{S}$ were instead written $\frac{S}{S^H}$ and $\frac{S}{S^{\text{He}}}$, respectively.

We note that the values of $S_T$ and $S_P$ are important because they determine the adiabatic gradient $\nabla_{\text{ad}}$

$$\nabla_{\text{ad}} = \left.\frac{\partial \log T}{\partial \log P}\right|_S = -\frac{S_P}{S_T}. \quad (9)$$

Therefore, the errors in the EOS propagated in the measurement of the specific heat at constant pressure and more strongly in the adiabatic gradient. When comparing the values of $\nabla_{\text{ad}}$ for both the correct and the incorrect table for the EOS, we found that the correct values are smaller, and that the average difference can be about 25% for the conditions in the interior of the protoplanet. Hence, in our previous study, $\nabla_{\text{ad}}$ was overestimated. According to the Schwarzschild criterion, convection starts when $\nabla > \nabla_{\text{ad}}$. As a consequence, the development of extended convective regions was delayed, preventing an earlier contraction of the envelope. Therefore, the later occurrence of the runaway accretion of gas lengthened the formation time, allowing for a higher accretion rate of solids that produced very massive cores.

## 3. Results

Here we show the results obtained after replacement of the EOS. The initial model consisted of a solid embryo of $5 \times 10^{-3}$ $M_\oplus$ placed at the current position of Jupiter (5.2 AU). We assumed that the core grows following the oligarchic regime, accreting planetesimals of the same size and density ($\rho_p = 1.5$ g cm$^{-3}$). We considered five different masses for the protoplanetary disc, from 6 to 10 times the MMSN, and two radii for the planetesimals: 10 and 100 km. As can be seen from Tables 1 and 2, the impact of the EOS on the entire formation process is significant: formation times are shorter and core masses are notably smaller when the corrected formulae are used. In agreement with previous results, when the nebula mass increases, formation times become shorter and core masses higher; however, the increase in the core mass is not as significant as in calculations performed using the previous EOS. Also, when accreted planetesimals are smaller, formation times are notably shorter and cores are more massive. This agrees qualitatively with our previous results. We define the cross–over mass ($M_{\text{cross}}$) to be the mass of the core when the protoplanet's total mass contains identical amounts of gas and solids, and the cross–over time ($t_{\text{cross}}$) to be the elapsed time when the cross–over mass is reached. If we compare Tables 1 and 2, for the same disc mass, we see that the cross–over time when accreted planetesimals have a radius of 10 km is, on average, 3 times shorter than when planetesimals have a radius of 100 km, and the mass of the core is $\sim 25\%$ higher. The total formation time does not differ much from the cross–over time, especially for the more massive nebulae. On the other hand, core masses at the end of the simulation can be several Earth masses higher than the corresponding cross–over mass, especially if the isolation mass was not reached.

**Table 1.** Results of the formation of a Jupiter–like planet, where the accreted planetesimals have a radius of 100 km[a].

| Disc density [MMSN] | Previous results | | New results | |
|---|---|---|---|---|
| | Cross–over time [My] | Cross–over mass [$M_\oplus$] | Cross–over time [My] | Cross–over mass [$M_\oplus$] |
| 6  | 11.40 | 44.70 | 7.20 | 25.50 |
| 7  | 8.75  | 55.40 | 5.58 | 27.90 |
| 8  | 7.00  | 66.60 | 4.60 | 29.30 |
| 9  | 5.70  | 77.60 | 3.78 | 30.90 |
| 10 | 4.80  | 88.50 | 3.20 | 31.90 |

[a] The first column indicates the disc density in terms of the MMSN, the second and third columns show the results corresponding to Paper 1, and the last two columns list the updated results for exactly the same conditions but with the corrected interpolation formulae for the SCVH's EOS for a mixture of hydrogen and helium.

**Table 2.** The same as in Table 1 but considering that the radius of the accreted planetesimals is 10 km.

| Disc density [MMSN] | Previous results | | New results | |
|---|---|---|---|---|
| | Cross–over time [My] | Cross–over mass [$M_\oplus$] | Cross–over time [My] | Cross–over mass [$M_\oplus$] |
| 6  | 4.95 | 46.70 | 2.43 | 31.80 |
| 7  | 3.53 | 58.70 | 1.97 | 34.80 |
| 8  | 2.65 | 71.40 | 1.49 | 36.80 |
| 9  | 2.00 | 84.85 | 1.32 | 38.60 |
| 10 | 1.65 | 99.00 | 1.10 | 39.90 |

**Table 3.** Input parameters for cases J1, J3 and J7 of P96[a].

| Case | $\Sigma$ [g cm$^{-2}$] | $\rho$ [g cm$^{-3}$] | $T$ [K] | $r_p$ [km] | $t_{\rm cross}$ [My] | $M_{\rm cross}$ [$M_\oplus$] |
|---|---|---|---|---|---|---|
| J1 | 10 | $5 \times 10^{-11}$ | 150 | 100 | 7.58 | 16.17 |
| J3 | 15 | $5 \times 10^{-11}$ | 150 | 100 | 1.51 | 29.61 |
| J7 | 10 | $5 \times 10^{-11}$ | 150 | 1   | 6.94 | 16.18 |

[a] Here $\Sigma$ is the solids surface density of the disc, $\rho$ is the nebular density, $T$ is the nebular temperature (all at the position of Jupiter) and $r_p$ is the radius of accreted planetesimals. The results of P96 for $t_{\rm cross}$ and $M_{\rm cross}$ are also listed.

In Paper 1, we presented, for comparison, a simulation of the formation of Jupiter with the same initial and boundary conditions as in Case J3 of P96 (see Table 3). The main difference between P96's calculations and ours is the solids accretion regime prescribed, which in their case is faster than the oligarchic regime considered here. The recalculated cross–over time is 13.2 My and the cross–over mass, 21.4 $M_\oplus$ (previously $t_{\rm cross} \sim 19$ My and $M_{\rm cross} \sim 29$ $M_\oplus$). Although $t_{\rm cross}$ is now shorter, it is still one order of magnitude longer than in the simulation of P96, where $t_{\rm cross} \simeq 1.5$ Myr. This clearly illustrates the impact of the core accretion rate on the formation timescale, because the time required to form an oligarchically growing core represents a significant fraction of the total formation time of the planet. This is a consequence of the gravitational perturbations produced by the embryo on the surrounding planetesimals, as it increases their relative velocities preventing mutual accretion. Planetesimals then become less effectively captured by the protoplanet. If we consider again the same conditions as for Case J3 but reduce the radius assigned to the incoming planetesimals we find that: for $r_p = 10$ km, $t_{\rm cross}$ is 3.7 My and $M_{\rm cross}$ is 25.5 $M_\oplus$, and for $r_p = 1$ km, $t_{\rm cross}$ is 1.4 My and $M_{\rm cross}$ is 28.8 $M_\oplus$ (Fig. 1). We note that the formation time is strongly dependent on planetesimals radii, being shorter when the considered radius is smaller. From this example, we clearly see that by just reducing the planetesimal's radius from 100 to 1 km (all of which are plausible values), the formation time becomes one order of magnitude shorter.

We also performed, for completeness, simulations with the same conditions of the nominal Case J1 of P96, considering the oligarchic growth for the core. The difference between Case J1 and Case J3 is only the initial surface density of the protoplanetary disc at the position of Jupiter; being lower in Case J1 (see Table 3. We note that these nebula conditions are compatible with a 3 MMSN). Our results for Case J1 are $t_{\rm cross} \simeq 23.25$ My and $M_{\rm cross} \simeq 15.75$ $M_\oplus$, while in P96 $t_{\rm cross} \simeq 7.6$ My and $M_{\rm cross} \simeq 16.17$ $M_\oplus$. Although the cross–over mass is almost the same in both simulations, the time of cross–over in our case is three times longer due to the adoption of a slower solids accretion regime. If we considered that the radii of the accreted planetesimals are 10 km, instead of 100 km, we found that $t_{\rm cross} \simeq 9.75$ My, while the cross–over mass remains practically unchanged. However, a very different situation is evident for $r_p = 1$ km: in this case, we do not find a stable solution (Fig. 2). P96 calculated this simulation (case J7) and their results are stable, being $t_{\rm cross} \simeq 6.94$ My and $M_{\rm cross} \simeq 16.18$ $M_\oplus$. Unstable solutions as presented here were first found in Benvenuto et al. (2007), but in their article the incorrectly interpolated EOS of SCVH was used. A detailed analysis of this simulation confirms that the conclusions of Benvenuto et al. (2007) also remain valid.

## 4. Discussion and conclusions

In this article, we have presented calculations of *in situ* giant–planet formation in the framework of the core–instability hypothesis. Most of these simulations were presented previously in Paper 1. The results reported here were obtained by correcting typos in the interpolation formulae for a mixture of hydrogen and helium given in SCVH. The new calculations have discovered both lower core masses and shorter formation times. This result agrees that the affected quantities of the EOS were the entropy and its derivatives, errors that translated into larger adiabatic gradients that, in turn, delayed the formation of the convective regions in the protoplanet's interior. It is important to mention that the most badly affected results were those corresponding to high density discs. For example, for a "low" density nebula, as in Case J3, the previous value of the cross–over mass was overestimated by 35%, while for the 10 MMSN the overestimation was 177%. Although these typos affected significantly the values of $t_{\rm cross}$ and $M_{\rm cross}$, they did not change the main conclusions of our previous paper.

Adopting the oligarchic growth regime for the core strongly slows down the entire formation process of giant planets compared to cases where a rapid core accretion rate is considered (e.g., P96). Moreover, from the shape of the curve of cumulative mass versus time shown in Fig. 1, we can only distinguish a first long phase where the core forms and simultaneously accretes gas, and a short runaway phase after which the cross–over mass is reached, where the protoplanet accretes most of its gaseous mass in a negligible time. This is another difference with the results of P96, where three phases can be distinguished in the mass–accretion process. It is clear from Fig. 1 that "phase 2" is absent in our simulations. This result, which was also pointed out in Paper 1, agrees with the semi–analytical studies of Shiraishi & Ida (2008).

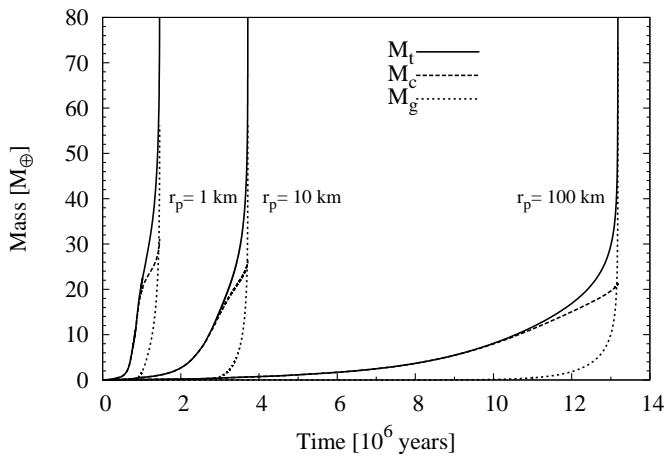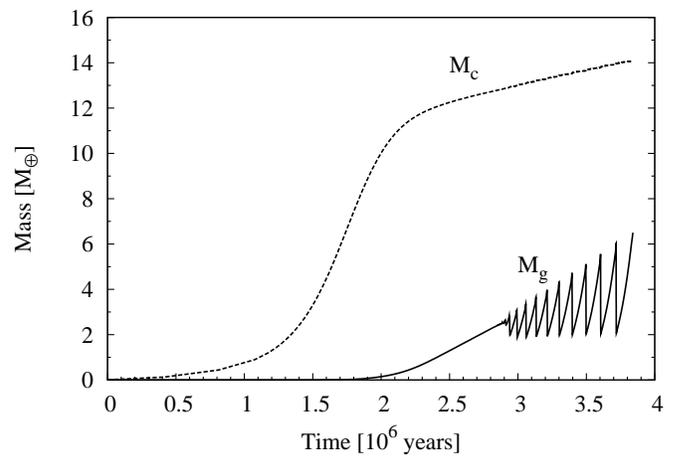

**Fig. 1.** Cumulative masses as a function of time for the formation of a Jupiter–like object. With a solid line, we plot the total mass, with a dashed line the core mass, and with a dotted line the envelope mass. Nebular conditions are the same as in Case J3 of P96. We considered three cases: one where the radius of accreted planetesimals is as in the original case ($r_p = 100$ km) and the other two for $r_p = 10$ km and $r_p = 1$ km. Clearly, the formation time is strongly dependent on the planetesimal's radius, while the mass of the core at the moment of cross–over remains approximately the same.

**Fig. 2.** An unstable solution: for the nominal case of P96 but with a planetesimal radius of 1 km. The evolution of the mass of the core (envelope) is plotted in dashed (solid) line.

We explored the dependence of the formation process on both the size of the accreted planetesimals and the surface density of the protoplanetary disc. From our simulations, we can conclude that reducing the size of accreted planetesimals is more effective to shortening the formation timescale than considering very massive nebulae. For example, in calculating the formation of Jupiter for a 10 MMSN (the most massive one considered here), the formation time was found to be 3.2 My for planetesimals of 100 km, while for a disc of 6 MMSN and planetesimals of 10 km the formation time is 2.43 My, the mass of the core in both cases being nearly the same (Tables 1 and 2).

Whether planetesimals that populate the feeding zone of an embryo are small or large remain unclear. Planetesimal formation is a problem that has yet to be solved. According to Ida et al. (2008), accretion and fragmentation among planetesimals are two competing processes. Moreover, the presence of turbulence in the disc can impede the accretion. While calculations have shown that it is possible to form protoplanetary embryos of Ceres mass on very short timescales (Johansen et al. 2007), giant planets require the presence of small planetesimals to complete the formation within the disc lifetime. Moreover, numerical simulations showed that during the oligarchic growth, the protoplanet is surrounded by smaller planetesimals (Kokubo & Ida 2000). At the moment, the average size of the planetesimals during the formation of the giant planets is unclear and further work is needed.

Finally, despite the typos in the EOS, the unstable solutions found by Benvenuto et al. (2007) remain. The simulation presented in this article (Fig. 2) is representative of the oscillatory instability that can affect the entire mass of the envelope if the isolation mass is reached before a significant amount of gas is bound to the protoplanet.

We note that ablation of accreted planetesimals has not been considered here. Ablation enlarges the cross-section of the protoplanet accelerating the formation process (Benvenuto & Brunini 2008). However, it also modifies both the opacity and the EOS of the envelope, which on the other hand, could delay the formation. Small planetesimals could also completely disintegrate before they reach the core, reducing its final mass.